\documentstyle[aps]{revtex}
\begin{document}
\title{Vortex states in binary mixture of
Bose-Einstein condensates}
\author{ S. T. Chui$^a$, V. N. Ryzhov$^b$ and E. E.
Tareyeva$^b$}
\address{a: Bartol Research Institute, University of
Delaware, Newark, DE
19716\\b:Institute for High Pressure Physics, Russian
Academy of
Sciences, 142 092 Troitsk, Moscow region, Russia}
\maketitle

\begin{abstract}
The vortex configurations in the Bose-Einstein condensate
of the mixture of two different spin states $|F=1,m_f=-
1>$ and $|2,1>$ of $^{87}Rb$ atoms corresponding to the
recent experiments by Matthews {\it et. al.} (Phys. Rev.
Lett. {\bf 83}, 2498 (1999)) are considered in the
framework of the Thomas-Fermi approximation as functions
of $N_2/N_1$, where $N_1$ is the number of atoms in the
state $|1,-1>$ and $N_2$ - in the state $|2,1>$. It is
shown that for nonrotating condensates
the configuration with the $|1,-1>$ fluid forming the
shell about the $|2,1>$ fluid (configuration "a") has
lower energy than the opposite configuration
(configuration "b") for all values of $N_2/N_1$. When the
$|1,-1>$ fluid has net angular momentum and forms an
equatorial ring around the resting central condensate
$|2,1>$, the total energy of the system
is higher than the ground energy, but the configuration
"a" has lower energy than the configuration "b" for all
$N_2/N_1$. On the other hand, when the $|2>$ fluid has
the net angular momentum, for the lowest value of the
angular momentum $\hbar l$ ($l=1$) there is the range of
the ratio $N_2/N_1$ where the configuration "b" has lower
energy than the configuration "a". For higher values of
the angular momentum the configuration "b" is stable for
all values of $N_2/N_1$.
\end{abstract}


\twocolumn

The realization of Bose-Einstein Condensation (BEC)
in dilute atomic gases offers new opportunities for
studying quantum degenerate fluids \cite{[1]}.
These condensates which contain thousands
of atoms confined to microscale clouds have
similarities to superfluidity and laser, and provide new
testing ground for many body physics.

Bose-Einstein condensates are quantum fluids in which
many particles
have the same quantum state. Intrinsic property of the
interacting
Bose-Einstein condensate is superfluidity. Bulk
superfluids are
distinguished from normal
fluids by their ability to support dissipationless
flow. Superfluidity of $^4He$ atoms has been widely
studied,
however, only recently evidence for critical velocity - a
key
characteristic of superfluids - was observed in a Bose
condensate
of sodium atoms \cite{crit_vel}.

Superfluidity is closely related to the existence of
stable quantized
vortices which are well known for superfluid $^4He$.
Vortices can be created in superfluid $^4He$ by cooling a
rotating
container with helium through the superfluid transition.
Quite recently the vortices in one-component Bose-
Einstein
condensate were generated by simple "stirring" of
$^{87}Rb$ atoms
condensate \cite{stir2000}. Vortex states in binary
mixture of
dilute Bose-Einstein condensates have been
created \cite{bose_vor} using the method proposed by
Williams
and Holland \cite{wilhol}. Their new scheme exploits the
possibility of simultaneously trapping otherwise
identical atoms
of $^{87}Rb$ in two different "hyperfine" spin states
$|1>$ and
$|2>$. ($|1>$ and $|2>$ denote $|F=1, m_f=-1>$ and
$|2,1>$ spin
states of $^{87}Rb$ atoms respectively). Matthews {\it et
al.}
\cite{bose_vor} applied a microwave to the condensate and
focused
a laser beam at various points around its circumference,
splitting
the atoms into two hyperfine states. The result is a
double
condensate with one condensate at rest in the center of
the cloud
and the other in a unit (or multiple) vortex state.
Controlling the temporal and
spatial dependence of the microwave-induced conversion of
$|1>$ into $|2>$
(and vice versa), one can directly create a $|2>$ (or
$|1>$) state wave
function having a wide variety of shapes out of a $|1>$
(or $|2>$) ground-state wave function \cite{bose_vor} or
different
numbers of atoms in each state.

In \cite{bose_vor} two types of vortices have been
formed. Matthews
{\it et al.} put the initial condensate into either the
$|1>$ or
$|2>$ state and then made a vortex in the $|2>$ or $|1>$
state.
They considered the time evolution of the vortices to
study their
dynamics and stability and showed that the dynamics of
the $|1>$
state vortices is different from the dynamics of the
$|2>$ state
vortices due to different scattering lengths.

The simultaneously
trapped resting condensates consisting of the $^{87}Rb$
atoms in the
$|1,-1>$ and $|2,1>$ spin states have been considered
experimentally in \cite{dhall}. In this case the
intraspecies
and interspecies scattering lengths
denoted correspondingly as $a_{11}, a_{12},, a_{22}$
are in the proportion $a_{11} :a_{12} :a_{22}
=1.03:1:0.97$
with the average of the three being $55(3)\AA$. It
has been observed that the atoms with the larger
scattering
length $a_{11}$ in the state $|1>$ form a lower-density
shell
about the atoms with the smaller scattering length
$a_{22}$ \cite{dhall}.
In \cite{bose_vor} it has been shown that when the $|1>$
fluid
forms an equatorial ring around the resting central fluid
this
configuration is rather stable. Conversely, the $|2>$
vortex ring
sinks in towards the trap center and breaks up.

The aim of this article is to study the relative
stability of different
vortex configurations as a function of the ratio
$N_2/N_1$, where
$N_1$ is the number of atoms in the state $|1>$ and $N_2$
- in the state
$|2>$. It is shown that for nonrotating phase-separated
condensates
the configuration with the $|1>$ fluid forming the shell
about the $|2>$
fluid (configuration "a") has lower energy than the
opposite configuration
(configuration "b") for all values of $N_2/N_1$. When the
$|1>$ fluid has
net angular momentum, the total energy of the system in
nonrotating
container is higher than the ground energy, but
configuration "a" has
lower energy than the configuration "b" for all
$N_2/N_1$. On the other
hand, when the $|2>$ fluid has the net angular momentum,
for the lowest
value of the angular momentum $\hbar l$ ($l=1$) there is
the range of
the ratio $N_2/N_1$ where the configuration "b" has lower
energy than
the configuration "a". For higher values of the angular
momentum the
configuration "b" is stable for all values of $N_2/N_1$.
It should be
noted that the experiment of Matthews {\it et al.}
\cite{bose_vor}
corresponds to the case where configuration "b" is
unstable
($N_2/N_1 \approx 1, l=1$). In our calculations we use
the parameters
corresponding to the experiments by Matthews {\it et al.}
\cite{bose_vor}:
approximately $8\times 10^5$ atom in a spherically
symmetric potential
with oscillation frequencies of $7.8 Hz$ in the radial
and axial directions
for both spin states. Let us describe our results in
detail.

	The modern theoretical description of dilute BEC
was originated by the seminal Bogoliubov's
1947 paper where he showed that weak repulsive
interaction qualitatively changes the excitation spectra
from quadratic free particle form to
a linear phonon-like structure. To describe the trapped
condensates
at $T=0$ one can use the Gross-Pitaevskii (GP)
(nonlinear Schrodinger) equation for the condensate wave
function \cite{[2]}. This equation appears as
the generalization of the Bogoliubov theory for the
inhomogeneous phase. It was widely used to discuss
the ground state properties and collective excitations in
one-component BEC \cite{[3],[4],[5],[6]}.

	In order to derive analytic results,
some approximations must be used. A commonly used one
is the Thomas-Fermi Approximation (TFA),
which ignores the kinetic energy terms. It has been shown
that in the case of one component condensates the TFA
results
agree well  with the numerical
calculations for large particle numbers,
except for a small region near the boundary of the
condensate
\cite{[4],[5]}. In fact, even for small numbers of
particles
TFA still usually gives qualitatively correct results.
In some situations
the TFA cannot be used to predict quantitative
features of the binary mixture of BEC. For example, the
TFA solution considerably underestimate
the degree of overlap between the condensates, however,
it provides an excellent
starting point of study.

Let us first consider the phase separation in binary
mixture
without rotation.

In the case of a two-species condensate, letting
$\psi_i({\bf r})$ $(i=1,2)$
be the wave function of species $i$ with particle number
$N_i$, we can
write two coupled nonlinear Schrodinger (Gross-
Pitaevskii) equations as:
\begin{eqnarray}
&-&\frac{\hbar^2}{2m_1}\nabla^2\psi_1({\bf r})+
\frac{1}{2}m_1\omega_1^2(x^2+y^2
+\lambda^2z^2)\psi_1({\bf r})-
\nonumber \\
&-&\mu_1\psi_1({\bf r})+
G_{11}|\psi_1({\bf r})|^2\psi_1({\bf
r})+G_{12}|\psi_2({\bf r})|^2
\psi_1({\bf r})=0\label{1}\\
&-&\frac{\hbar^2}{2m_2}\nabla^2\psi_2({\bf
r})+\frac{1}{2}m_2\omega_2^2(x^2+y^2
+\lambda^2z^2)\psi_2({\bf r})-
\nonumber \\
&-&\mu_2\psi_2({\bf r})+
G_{22}|\psi_2({\bf r})|^2\psi_2({\bf
r})+G_{12}|\psi_1({\bf r})|^2
\psi_2({\bf r})=0\label{2}
\end{eqnarray}

Equations (\ref{1}) and (\ref{2}) were obtained by
minimization of the
energy functional of the trapped bosons of masses $m_1$
and $m_2$ given by:
\begin{eqnarray}
&&E(\psi_1,\psi_2)=
=\int\,d^3 r\left[
\frac{\hbar^2}{2m_1}|\nabla\psi_1({\bf
r})|^2+\right.
\nonumber\\
&+&\left.\frac{1}{2}m_1\omega_1^2(x^2+y^2
+\lambda^2z^2)|\psi_1({\bf r})|^2+\right. \nonumber\\
&+&\frac{\hbar^2}{2m_2}|\nabla\psi_2({\bf
r})|^2+\frac{1}{2}m_2\omega_2^2(x^2+y^2
+\lambda^2z^2)|\psi_2({\bf r})|^2+ \nonumber\\
&+&\left.\frac{G_{11}}{2}|\psi_1({\bf
r})|^4+\frac{G_{22}}{2}|\psi_2({\bf r})|^4
+G_{12}|\psi_1({\bf r})|^2|\psi_2({\bf r})|^2 \right].
\label{3}
\end{eqnarray}

The chemical potentials $\mu_1$ and $\mu_2$ are
determined by the relations
$\int\,d^3 r|\psi_i|^2=N_i$. The trap potential is
approximated
by an effective three-dimensional harmonic-oscillator
potential
well, which is cylindrically symmetric about $z$ axis,
$\lambda$
being the ratio of angular frequencies in the axial
direction
$\omega_{zi}$ to that in the transverse direction
$\lambda=\omega_{zi}/\omega_i$.  The interaction
strengths, $G_{11}, G_{22}, G_{12}$ are
determined by the $s$-wave scattering lengths for binary
collisions of
like and unlike bosons: $G_{ii}=4\pi\hbar^2a_{ii}/m_i;
G_{12}=2\pi\hbar^2a_{12}
/m$, where $m^{-1}=m_1^{-1}+m_2^{-1}$.

Let us consider now the phase separation due to
interaction between two
condensates. In our case of $|1,-1>$ and $|2,1>$ we have
$\frac{1}{2}m_1\omega_1^2=\frac{1}{2}m_2\omega_2^2.$
We simplify the equations by using dimensionless
variables. Let us define
the length scale
$a_{\perp}=\left(\frac{\hbar}{m_1\omega_1}\right)^{1/2}$,
and define the dimensionless
variables
${\bf r}=a_{\perp}{\bf r}',
E=\hbar\omega_1 E',
\psi_i({\bf
r})=\sqrt{N_i/a_{\perp}^3}\psi_i'({\bf r}').$
The wave function $\psi_i'({\bf r}')$ is normalized to
$1$.
In terms
of these variables the Gross-Pitaevskii energy functional
takes the form:
\begin{eqnarray}
E'&=&\frac{1}{2}\int\,d^3r'\left[N_1|\nabla'\psi_1'|^2+
N_1(x'^2+y'^2+
\lambda^2z'^2)|\psi_1'|^2+
\right.\nonumber\\
&+&\left.
N_2\beta^2|\nabla'\psi_2'|^2+N_2(x'^2+y'^2+\lambda^2z'^2)
|\psi_2'|^2+ \right.
\nonumber\\
&+&\left.\frac{1}{2}N_1u_1|\psi_1'|^4+\frac{1}{2}N_2u_2\b
eta^2|\psi_2'|^4+
\right.\nonumber\\
&+&\left.
\frac{2\pi a_{12}}{a_{\perp}}\frac{m_1}{m}
N_1N_2|\psi_1'|^2|\psi_2'|^2\right].
\label{9}
\end{eqnarray}
Here $\beta^2=m_1/m_2=\omega_2^2/\omega_1^2$ and
$u_i=8\pi a_{ii}N_i/a_{\perp}$.
Eqs. (\ref{1}) and (\ref{2}) are rewritten as:
\begin{eqnarray}
&-&\nabla'^2\psi_1'+(x'^2+y'^2+\lambda^2z'^2)\psi_1'-\nonumber\\
&-&\mu_1'\psi_1'+
u_1|\psi_1'|^2\psi_1'+
+\frac{4\pi
a_{12}N_2}{a_{\perp}}\frac{m_1}{m}|\psi_2'|^2\psi_1'=0;
\label{10}\\
&-
&\beta^2\nabla'^2\psi_2'+(x'^2+y'^2+\lambda^2z'^2)\psi_2'
-\mu_2'\psi_2'+ \nonumber\\
&+&u_2\beta^2|\psi_2'|^2\psi_2'+
+\frac{4\pi
a_{12}N_1}{a_{\perp}}\frac{m_1}{m}|\psi_1'|^2\psi_2'=0;
\label{11}
\end{eqnarray}
where $\mu_i'=2\mu_i/\hbar\omega_1$.

In the TFA, Eqs. (\ref{9}), (\ref{10})
and (\ref{11}) can be further simplified by omitting the
kinetic energy.
For TFA the phase segregated condensates do not overlap,
so we can neglect
the last terms in Eqs. (\ref{9}), (\ref{10}) and
(\ref{11}), obtaining from
(\ref{10}) and (\ref{11}), in different regions (to be
determined later), the simple algebraic equations:
\begin{eqnarray}
|\psi_1'({\bf r}')|^2&=&\frac{1}{u_1}\left(\mu_1'-
(\rho'^2+\lambda^2z'^2)\right);
\label{12} \\
|\psi_2'({\bf r}')|^2&=&\frac{1}{u_2\beta^2}
\left(\mu_2'-(\rho'^2+\lambda^2z'^2)\right). \label{13}
\end{eqnarray}
Here  $\rho'^2=x'^2+y'^2$. From
Eqs. (\ref{12}) and (\ref{13}) one can see that the
condensate density has
the ellipsoidal form.

In the case of phase separation, the energy of the system
can be written
in the form
\begin{equation}
E=E_1+E_2, \label{14}
\end{equation}
where
\begin{eqnarray}
E_1&=&\frac{1}{2}\hbar\omega_1 N_1\left[\mu_1'-
\frac{1}{2}u_1\int\,d^3r'
|\psi_1'|^4\right], \label{15}\\
E_2&=&\frac{1}{2}\hbar\omega_1 N_2\left[\mu_2'-
\frac{1}{2}u_2\beta^2
\int\,d^3r' |\psi_2'|^4\right]. \label{16}
\end{eqnarray}
In order to obtain Eqs. (\ref{15})-(\ref{16}),
Eqs. (\ref{12})-(\ref{13}) have been used.

To investigate the phase separation in the mixture we
first
suppose that the condensate 1 atoms form an ellipsoidal
shell about
the condensate 2 atoms (configuration "a").
To determine the position of the boundary between
the condensates, we use the condition of thermodynamic
equilibrium
\cite{landau}: the pressures exerted by both condensates
must be equal:
\begin{equation}
P_1=P_2. \label{17}
\end{equation}
Pressure is given by \cite{pitaevskii}:
\begin{equation}
P_i=\frac{G_{ii}}{2}|\psi_i|^4. \label{18}
\end{equation}

Condensate 2 has the form of the ellipsoid with long
semiaxis $q$:
\begin{equation}
\rho'^2+\lambda^2z'^2=q^2. \label{19}
\end{equation}

From Eqs. (\ref{12})-(\ref{13}) and (\ref{17})-(\ref{19})
one has
the equation for $q$:
\begin{equation}
\mu_1'-q^2=\kappa \mu_2'-\kappa q^2, \label{20}
\end{equation}
where $\kappa=\sqrt{(a_{11}m_2)/(a_{22}m_1)}$.

Chemical potentials $\mu_1'$ and $\mu_2'$ can be obtained
using the
normalization conditions
$\int\,d^3r'|\psi_1'|^2=\int\,d^3r'|\psi_2'|^2=1$
and are given by:

\begin{equation}
\mu_1'=\frac{\mu_1^0}{\left(1-
\frac{5}{2}q'^3+\frac{3}{2}q'^5\right)^{2/5}},
\label{21}
\end{equation}
\begin{equation}
\mu_2'=\frac{3}{(\mu_1')^{3/2}q'^3}\left(\frac{2\beta^2(\mu_2^0)^{5/2}}{15}+
\frac{(\mu_1')^{5/2}q'^5}{5}\right), \label{22}
\end{equation}
where  $q=\sqrt{\mu_1'}q'$ and
\begin{equation}
\mu_i^0=\left(\frac{15\lambda u_i}{8\pi}\right)^{2/5}.
\label{23}
\end{equation}

From equations (\ref{21})-(\ref{23}) one can determine
the chemical
potentials $\mu_1'$ and $\mu_2'$ and the semiaxis of the
phase boundary ellipsoid $q$ as a function of $N_1$ and
$N_2$.
The energy of the configuration "a" $E_a=E_{a1}+E_{a2}$
is given by:
\begin{eqnarray}
E_{a1}&=&\frac{1}{2}\hbar \omega_1 N_1 \left\{\mu_1'-
\frac{15}{4}
\frac{(\mu_1')^{7/2}}{(\mu_1^0)^{5/2}}
\times\right.\nonumber\\
&\times&\left.\left[\frac{8}{105}
-
\left(\frac{q'^3}{3}-
\frac{2}{5}q'^5+\frac{q'^7}{7}\right)\right]\right\},
\label{24} \\
E_{a2}&=&\frac{1}{2}\hbar \omega_1 N_2 \left\{\mu_2'-
\frac{15}{4}
\frac{(\mu_1')^{3/2}}{\beta^2(\mu_2^0)^{5/2}}
\times\right.\nonumber\\
&\times&\left.
\left({\mu_2'}^2
\frac{q'^3}{3} -
2\mu_2'\mu_1'\frac{q'^5}{5}+\mu_1'^2\frac{q'^7}{7}\right)
\right\}.
\label{25}
\end{eqnarray}

Let us now consider the opposite case when
the condensate $|2>$ atoms form an ellipsoidal shell
about the condensate $|1>$
atoms (configuration "b"). In this case Eqs. (\ref{20})-
(\ref{25})
can be rewritten in the form:
\begin{equation}
\mu_1''-q_1^2=\kappa(\mu_2''-q_1^2),\label{26}
\end{equation}
\begin{equation}
(\mu_2'')^{5/2}=\frac{\beta^2(\mu_2^0)^{5/2}}{1-
\frac{5}{2}q_1'^3+
\frac{3}{2}q_1'^5},\label{27}
\end{equation}
\begin{equation}
\frac{15}{2}\frac{(\mu_2'')^{3/2}}{(\mu_1^0)^{5/2}}\left(
\frac{\mu_1''q_1'^3}{3}-\frac{\mu_2''q_1'^5}{5}\right)=1,
\label{28}
\end{equation}
\begin{eqnarray}
E_b&=&E_{b1}+E_{b2}, \nonumber\\
E_{b1}&=&\frac{1}{2}\hbar\omega_1 N_1 \left\{\mu_1''-
\frac{15}{4}
\frac{({\mu_2}'')^{3/2}}{(\mu_1^0)^{5/2}}
\times\right.\nonumber\\
&\times&\left.
\left({\mu_1}''^2
\frac{q_1'^3}{3}-
2\mu_1''\mu_2''\frac{q_1'^5}{5}+\mu_2''^2\frac{q_1'^7}{7}
\right)\right\},
\label{29}\\
E_{b2}&=&\frac{1}{2}\hbar\omega_1 N_2\left\{\mu_2''-
\frac{15}{4}
\frac{(\mu_2'')^{7/2}}{\beta^2(\mu_2^0)^{5/2}}
\times\right.\nonumber\\
&\times&\left.
\left[\frac
{8}{105}-\left(
\frac{q_1'^3}{3}-
\frac{2}{5}q_1'^5+\frac{q_1'^7}{7}\right)\right]\right\}.
\label{30}
\end{eqnarray}
Here $\mu_1''$ and $\mu_2''$ are the chemical potentials
in the
configuration "b", $q_1=\sqrt{\mu_2''}q_1'$ is the long
semiaxis
of the boundary ellipsoid, $E_b$ is the energy of the
configuration "b".

To estimate which configuration is stable, one has to
compare $E_a$ and
$E_b$.

To evaluate $\Delta E=E_a-E_b$ in general case it is
worth first to estimate
the energy of the phase boundary which arises due to
gradient terms
omitted in TFA. The surface energy per unit area, the
surface tension,
is defined as $\sigma=E_s/S$, where $E_s$ is the surface
energy, and $S$
is the interface area. $\sigma$ may be written in the
form \cite{[13],[14]}:
\begin{eqnarray}
\sigma&=&\frac{\hbar\omega_1}{2\sqrt{2}a_{\perp}^2}\left(
\frac{a_{12}}{\sqrt{a_{11}a_{22}}}-
1\right)^{1/2}(u_1u_2N_1N_2)^{1/4}\times\nonumber\\
&\times&
|\psi_1'||\psi_2'|(N_1|\psi_1'|^2+N_2|\psi_2'|^2)^{1/2}.
\label{49}
\end{eqnarray}

Taking into account that the surface area of the
ellipsoid
with the semiaxis $a_{\perp}q$ has the form:
\begin{equation}
S=2\pi a_{\perp}^2
q^2\left(1+\frac{1}{\lambda\sqrt{\lambda^2-1}}
\log\frac{1}{\lambda-\sqrt{\lambda^2-1}}\right),
\label{50}
\end{equation}
one can estimate the contribution of the surface energy
$E_s=\sigma S$
to the total energy of each configuration. To be
specific, we will use
the parameters corresponding to the  experiments of
Matthews {\it et. al.} \cite{bose_vor} on $^{87}Rb$
atoms. In this case $m_1=m_2$, $a_{\perp}=3.88\times
10^{-4} cm$, $\lambda=1$,
$N=N_1+N_2=0.8\times10^6$ atoms. As mentioned in the
introduction,
$a_{11} :a_{12} :a_{22} =1.03:1:0.97$
with the average of the three being $55(3)\AA$.

In Fig. 1 we show
the energies of configurations "a" and "b"
(including the surface energy) $E_a/(\hbar \omega_1)$
(solid line)
and  $E_b/(\hbar \omega_1)$ (dashed line) as functions of
$\log_{10}(N_2/N_1)$.
One can see that $E_a$ is always lower than $E_b$.
This is consistent with the qualitative assertion and
experimental
observation that it is
energetically favorable for the atoms with the larger
scattering length
to form a lower-density shell about the atoms with the
smaller
scattering length \cite{[10],pu}.
It should be noted that
the surface energy is much smaller
than the interaction energy because the scattering
lengths $a_{ij}$
have very close values (see Eq. (\ref{49})).

Let us now consider the condensates with nonzero net
momentum.
For vortex excitation with angular momentum $\hbar l_j$ ,
the condensate
wave function is given by
\begin{equation}
\psi_{l_{j}}({\bf r})=|\psi_{l_{j}}({\bf
r})|e^{il_j\phi}. \label{58}
\end{equation}

After substituting the wave function for the vortex
excitation (\ref{58}) in
Eq. (\ref{14}),
the equation describes the ground state of bosons
in an effective confinement potential
$l_1^2\hbar^2/2m_1\rho^2+
l_2^2\hbar^2/2m_2\rho^2+V_1+V_2$, where
$V_i=m_i\omega_i(\rho^2+
\lambda^2 z^2)/2$ and $\rho^2=x^2+y^2$. So within the TFA
the density
of the vortex state has the form:
\begin{eqnarray}
|\psi_1'({\bf r}')|^2&=&\frac{1}{u_1}\left(\mu_1'(l_1)-
(\rho'^2+\lambda^2z'^2)
-\frac{l_1'^2}{\rho'^2}\right); \label{60}\\
|\psi_2'({\bf r}')|^2&=&\frac{1}{u_2\beta^2}
\left(\mu_2'(l_2)-(\rho'^2+\lambda^2z'^2)-
\frac{\beta^2l_2'^2}{\rho'^2}\right).
\label{61}
\end{eqnarray}
The important new qualitative feature of a vortex in the
TFA is the
appearance of a small hole of radius $\xi$,
$\xi_i^2\propto
l_i^2/\mu_i(l_i)$, but the remainder of the condensate
density
is essentially unchanged. The fractional change in the
chemical
potentials caused by the vortex $(\mu_i'(l_i)-
\mu_i')/\mu_i'$
can be shown to be small \cite{[3],[6]}, of the order of
$1/N^{4/5}$. In calculation of physical quantities
containing
condensate density it is sufficient to retain the no-
vortex
density and simply cut off any divergent radial integrals
at the
appropriate core sizes $\xi_1^2=l_1^2/\mu_1'$ or
$\xi_2^2=\beta^2l_2^2/\mu_2'$. Note that using the
unperturbed density for
calculation of the vortex properties corresponds to the
hydrodynamic limit.

In the case of the phase segregated condensate, one finds
from Eqs. (\ref{60}-\ref{61}) and (\ref{15}-\ref{16})
that the energy
change due to the presence of the vortices
$\Delta E(l_1,l_2)$ has the form:
\begin{eqnarray}
\Delta E&=&\Delta E_{N_1}+\Delta E_{N_2}=
\frac{1}{2}\hbar \omega_1 N_1 \int\,d_3 r'
\frac{l_1^2}{\rho'^2}|\psi_1'|^2 +\nonumber\\
&+&
\frac{1}{2}\hbar \omega_1 N_2 \int\,d_3 r'
\frac{l_2^2\beta^2}{\rho'^2}|\psi_2'|^2.
\label{62} \end{eqnarray} In
the hydrodynamic limit $\psi_i'$ is given by Eqs.
(\ref{12}) and
(\ref{13}).

Let us consider the configuration "a".
In the hydrodynamic limit the
location of the phase boundary is given by Eq.
(\ref{19}). From (\ref{62})
one has:
\begin{eqnarray}
\frac{\Delta E_{N_1}^a}{\frac{1}{2}\hbar
\omega_1N_1}&=&\frac{5l_1^2
(\mu_1')^{3/2}}{(\mu_1^0)^{5/2}}\left\{\left[\ln\frac{2\mu_1'}{l_1}-
\frac{4}{3}
\right]-
\frac{3}{2}q'
\times\right.\nonumber\\
&\times&\left.
\left[\left(1-\frac{1}{3}
q'^2\right)\ln\frac{2\mu_1'q'}{l_1}-
\left(1-\frac{q'^2}{9}\right)\right]\right\}
\label{63}\\
\frac{\Delta E_{N_2}^a}{\frac{1}{2}\hbar
\omega_1N_2}&=&\frac{15l_2^2
(\mu_1')^{1/2}q'}{2(\mu_2^0)^{5/2}}\left[\left(\mu_2'-
\frac{1}{3}\mu_1'q'^2
\right)
\times\right.\nonumber\\
&\times&\left.
\ln\frac{2\sqrt{\mu_1'\mu_2'}q'}{l_2\beta}-
\left(\mu_2'-
\frac{\mu_1'q'^2}{9}\right)\right]
\label{64}
\end{eqnarray}

The energy for the configuration "b" has the form (the
location of the phase
boundary is given by Eqs. (\ref{26}-\ref{28})):
\begin{eqnarray}
\frac{\Delta E_{N_2}^b}{\frac{1}{2}\hbar
\omega_1N_2}&=&\frac{5l_3^2
(\mu_2'')^{3/2}}{(\mu_2^0)^{5/2}}\left\{\left[\ln\frac{2\
mu_2''}{l_3\beta}
-\frac{4}{3} \right]-
\frac{3}{2}q_1'
\times\right.\nonumber\\
&\times&\left.
\left[\left(1-\frac{1}{3}
q_1'^2\right)\ln\frac{2\mu_2'q_2'}{l_3\beta}-
\left(1-\frac{q_1'^2}{9}\right)\right]\right\}
\label{63b}\\
\frac{\Delta E_{N_1}^b}{\frac{1}{2}\hbar
\omega_1N_1}&=&\frac{15l_4^2
(\mu_2'')^{1/2}q_1'}{2(\mu_1^0)^{5/2}}\left[\left(\mu_1''
-\frac{1}{3}\mu_2''q_1'^2
\right)
\times\right.\nonumber\\
&\times&\left.
\ln\frac{2\sqrt{\mu_1''\mu_2''}q_1'}{l_4}-
\left(\mu_1''-
\frac{\mu_2''q_1'^2}{9}\right)\right] \label{64b}
\end{eqnarray}

Using Eqs.(\ref{63}-\ref{64b}) one can compare the
energies of
the configurations for different values of the vortex
excitation
net angular momenta and different number of state $|1>$
and state $|2>$
atoms. The results of calculations for the case of
nonzero net
momenta are presented on the figures 2-7. The parameters
corresponding to the experiments by Matthews {\it et al.}
\cite{bose_vor}
were used.

The figures 2-4 show the situation with nonrotating
condensate $|2>$
atoms and a vortex of condensate $|1>$ atoms. In Fig.2 we
show the
rotational parts (\ref{63}-\ref{64b}) of the energies
of configurations "a" and "b"
 $\Delta E_a/(\hbar \omega_1)$ (solid line)
and  $\Delta E_b/(\hbar \omega_1)$ (dashed line) as
functions of
$\log_{10}(N_2/N_1)$ in the case when the condensate of
$|1>$ atoms has net angular momentum $l=1$ (lower curves)
and $l=2$ (upper curves). For the configuration "a" this
means the rotation of the outer shell and for the
configuration "b" -- of the inner part.

In Fig. 3 the corresponding total energy is shown, and
Fig. 4 shows the difference between $E_a$ and
$E_b$ for this case.
One can see that $E_a$ is always lower than $E_b$.

Let us consider now the case when the condensate $|2>$
atoms has the
net angular momentum $l=1$ or $l=2$. This case
corresponds to
the rotation of the outer shell for the configuration "b"
and of
the interior part of the condensate for the configuration
"a".
The rotational part is shown on the Fig. 5, the total
energy
-- on Fig. 6. The difference between the total energies
$E_a$ and $E_b$ shown in Fig. 7 presents the most
interesting of
our results. In the $l=2$ case the "b" configuration with
vortex
ring of condensate $|2>$ atoms in the outer shell is
stable in the
whole range of the relative concentrations $N_1$ and
$N_2$ while
in the case of $l=1$ vortex of condensate $|2>$ atoms
this configuration is stable only in the regions of low
$N_2/N_1$ or
$N_1/N_2$ concentrations. When the values $N_1$ and $N_2$
approach one another the condensate $|2>$ atoms ring
sinks in toward the trap center.

These results are in agreement with the experiment
\cite{bose_vor}  at approximately equal values
of $N_i$: when the condensate $|1>$ has a net angular
momentum, it
forms an equatorial ring around the central condensate
$|2>$.
Conversely, a condensate $|2>$ vortex forms a ring that
tends to
contract down into the condensate $|1>$.

It should be noted that the similar conclusions have been
made in Ref. \onlinecite{garcia}. Using the dynamic
stability analysis of the Gross-Pitaevskii equations, the
authors of Ref. \onlinecite{garcia} have investigated the
instability mechanism of the configuration "b" and
concluded that stabilization of this configuration can be
attained by the controlling the relative population of
both species.

To summarize, we have shown
that for nonrotating phase-separated condensates
the configuration with the $|1>$ fluid forming the shell
about the $|2>$
fluid (configuration "a") has lower energy than the
opposite configuration
(configuration "b") for all values of $N_2/N_1$. When the
$|1>$ fluid has
net angular momentum, the total energy of the system
is higher than the ground energy, but the configuration
"a" has
lower energy than the configuration "b" for all
$N_2/N_1$. On the other
hand, when the $|2>$ fluid has the net angular momentum,
for the lowest
value of the angular momentum $\hbar l$ ($l=1$) there is
the range of
the ratio $N_2/N_1$ where the configuration "b" has lower
energy than
the configuration "a". For higher values of the angular
momentum the
configuration "b" is stable for all values of $N_2/N_1$.

This work was supported in part by NATO Grant
No.PST.CLG.976038. V.N.R and E.E.T acknowledge the
financial support from the Russian Science Foundation
through the Grant No.98-02-16805. STC was partly
supported by NASA under grant No. CRG8-1427.
The authors are grateful to V.M. Perez-Garcia and J.J.
Garcia-Ripoll for drawing their attention to Ref.
\onlinecite{garcia}.

\begin{figure}
\caption{
The total energies of configurations "a" and "b"
$E_a/(\hbar \omega_1N)$ (solid line)
and  $E_b/(\hbar \omega_1N)$ (dashed line) for non
rotating condensate as functions of $\log_{10}(N_2/N_1)$.
}
\end{figure}
\begin{figure}
\caption{ The
rotational energies of configurations "a" and "b"
 $\Delta E_a/(\hbar \omega_1)$ (solid line)
and  $\Delta E_b/(\hbar \omega_1)$ (dashed line) as
functions of $\log_{10}(N_2/N_1)$ in the case when the
condensate of $|1>$ atoms has net angular momentum $l=1$
(lower curves) and $l=2$
(upper curves). }
\end{figure}
\begin{figure}
\caption{ The total energies of configurations "a" and
"b" $E_a/(\hbar \omega_1)$ (solid line)
and  $E_b/(\hbar \omega_1)$ (dashed line) as functions of
$\log_{10}(N_2/N_1)$ in the case when the condensate of
$|1>$ atoms has net angular momentum $l=1$ (lower curves)
and $l=2$
(upper curves). }
\end{figure}
\begin{figure}
\caption{ The difference of total energies for
configurations "a" and "b" $(E_a-E_b)/(\hbar \omega_1)$
as a function of $\log_{10}(N_2/N_1)$ in the case when
the condensate of $|1>$ atoms has net angular momentum
$l=1$ (solid line) and $l=2$ (dashed line).}
\end{figure}
\begin{figure}
\caption{ The
rotational energies of configurations "a" and "b"
 $\Delta E_a/(\hbar \omega_1)$ (solid line)
and  $\Delta E_b/(\hbar \omega_1)$ (dashed line) as
functions of $\log_{10}(N_2/N_1)$ in the case when the
condensate of $|2>$ atoms has net angular momentum $l=1$
(lower curves) and $l=2$ (upper curves). }
\end{figure}
\begin{figure}
\caption{ The total energies of configurations "a" and
"b" $E_a/(\hbar \omega_1)$ (solid line)
and  $E_b/(\hbar \omega_1)$ (dashed line) as functions of
$\log_{10}(N_2/N_1)$ in the case when the condensate of
$|2>$ atoms has net angular momentum $l=1$ (lower curves)
and $l=2$ (upper curves). }
\end{figure}
\begin{figure}
\caption{ The difference of total energies for
configurations "a" and "b" $(E_a-E_b)/(\hbar \omega_1)$
as a function of $\log_{10}(N_2/N_1)$ in the case when
the condensate of $|2>$ atoms has net angular momentum
$l=1$ (solid line) and $l=2$ (dashed line).}
\end{figure}

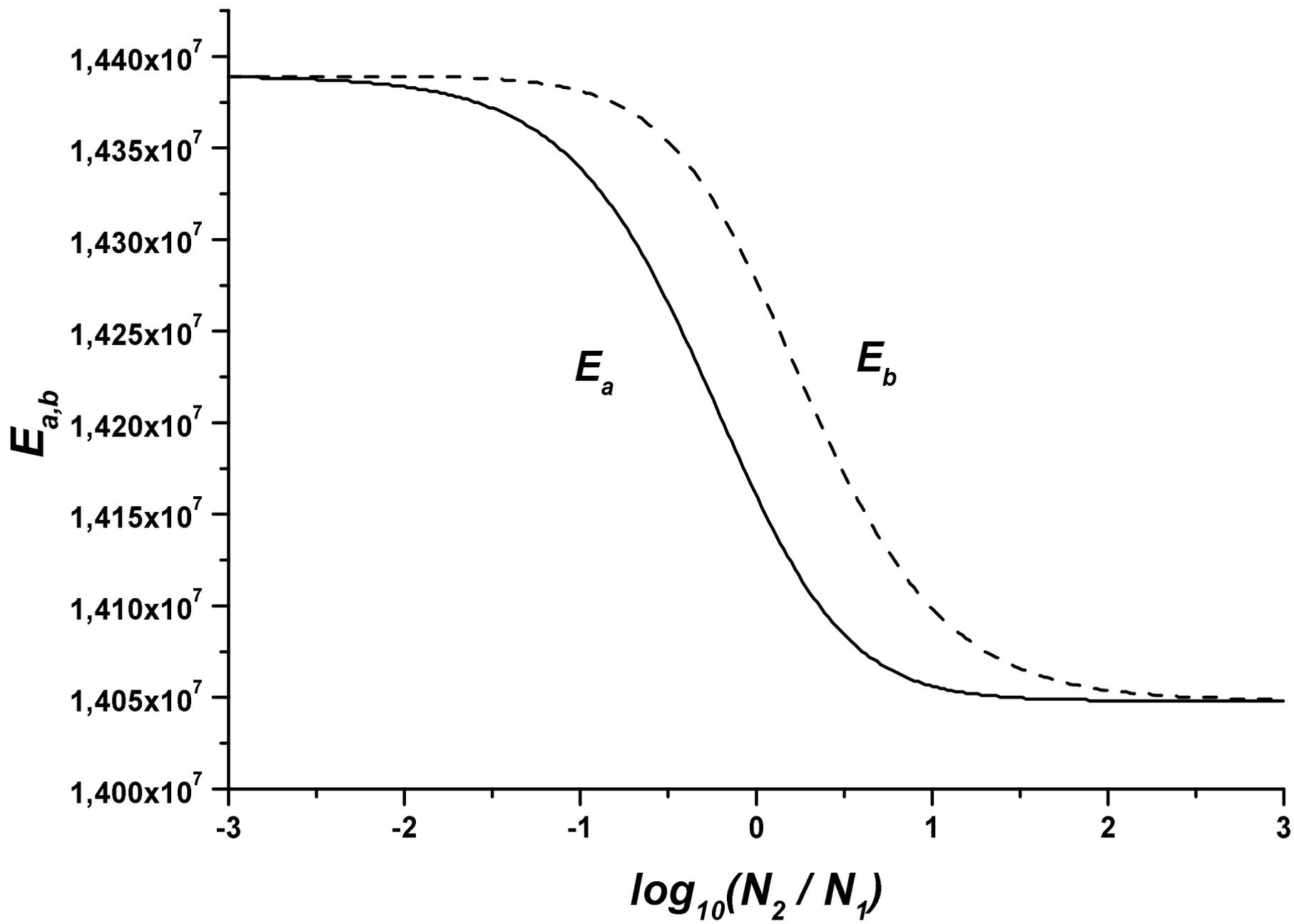
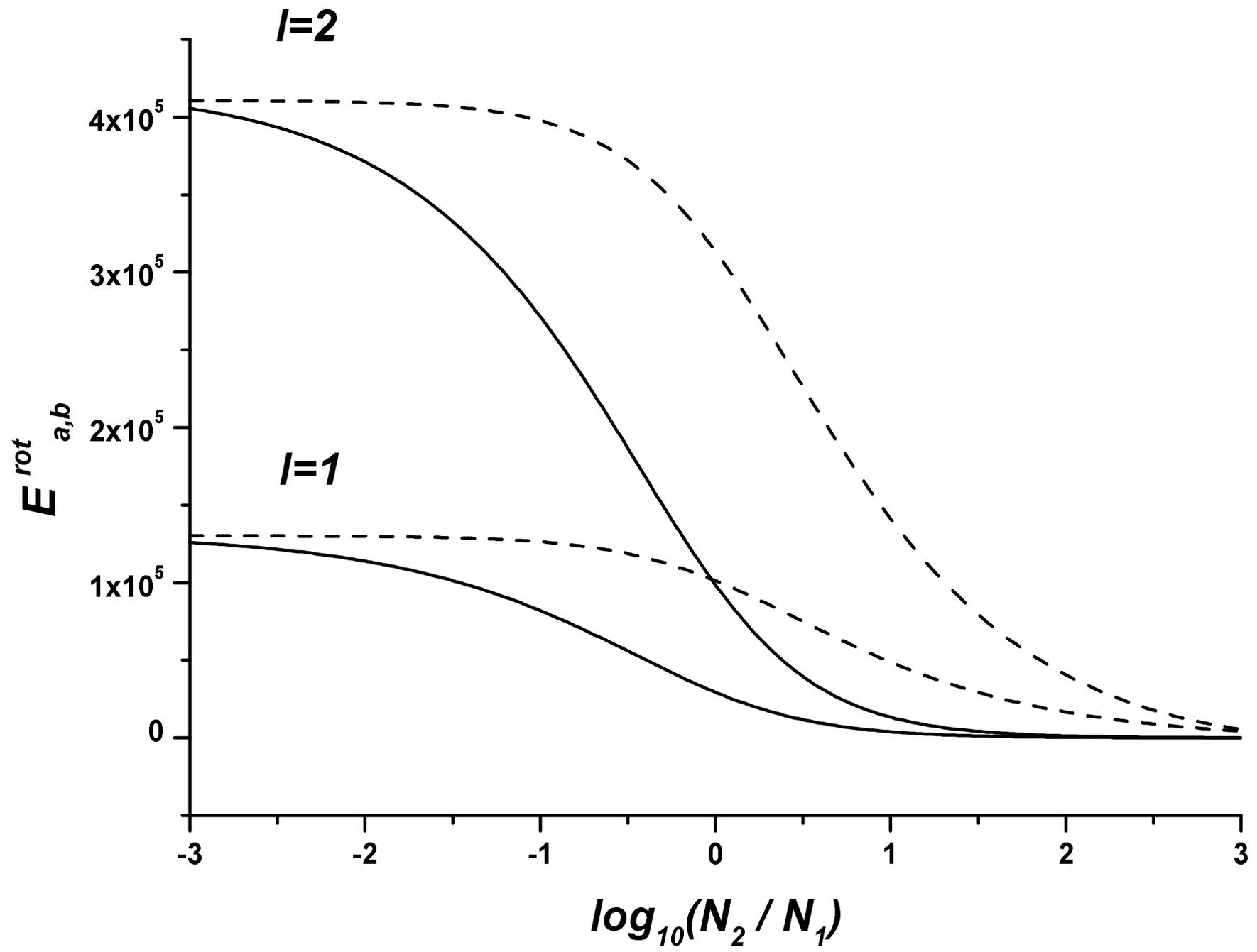
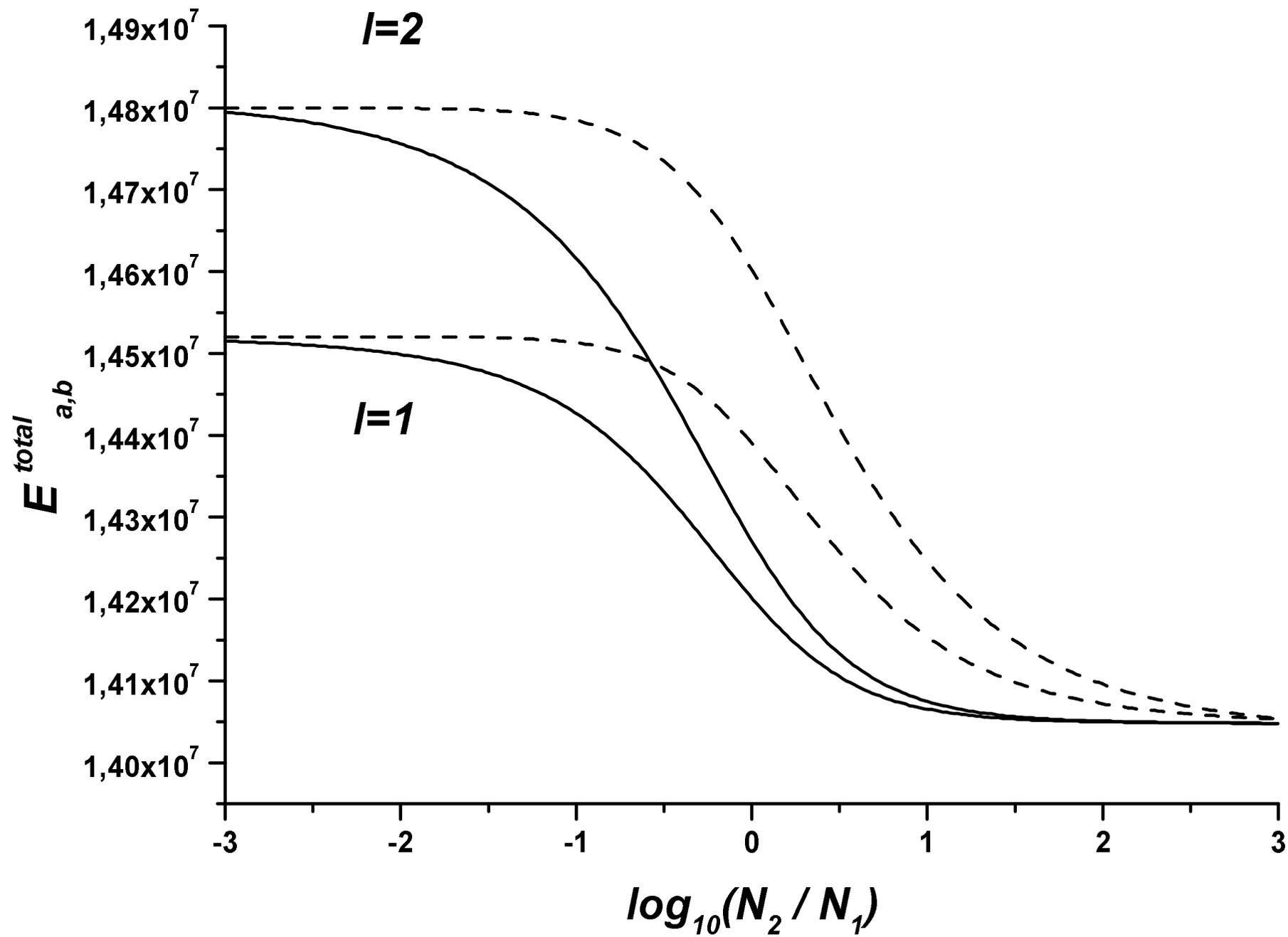
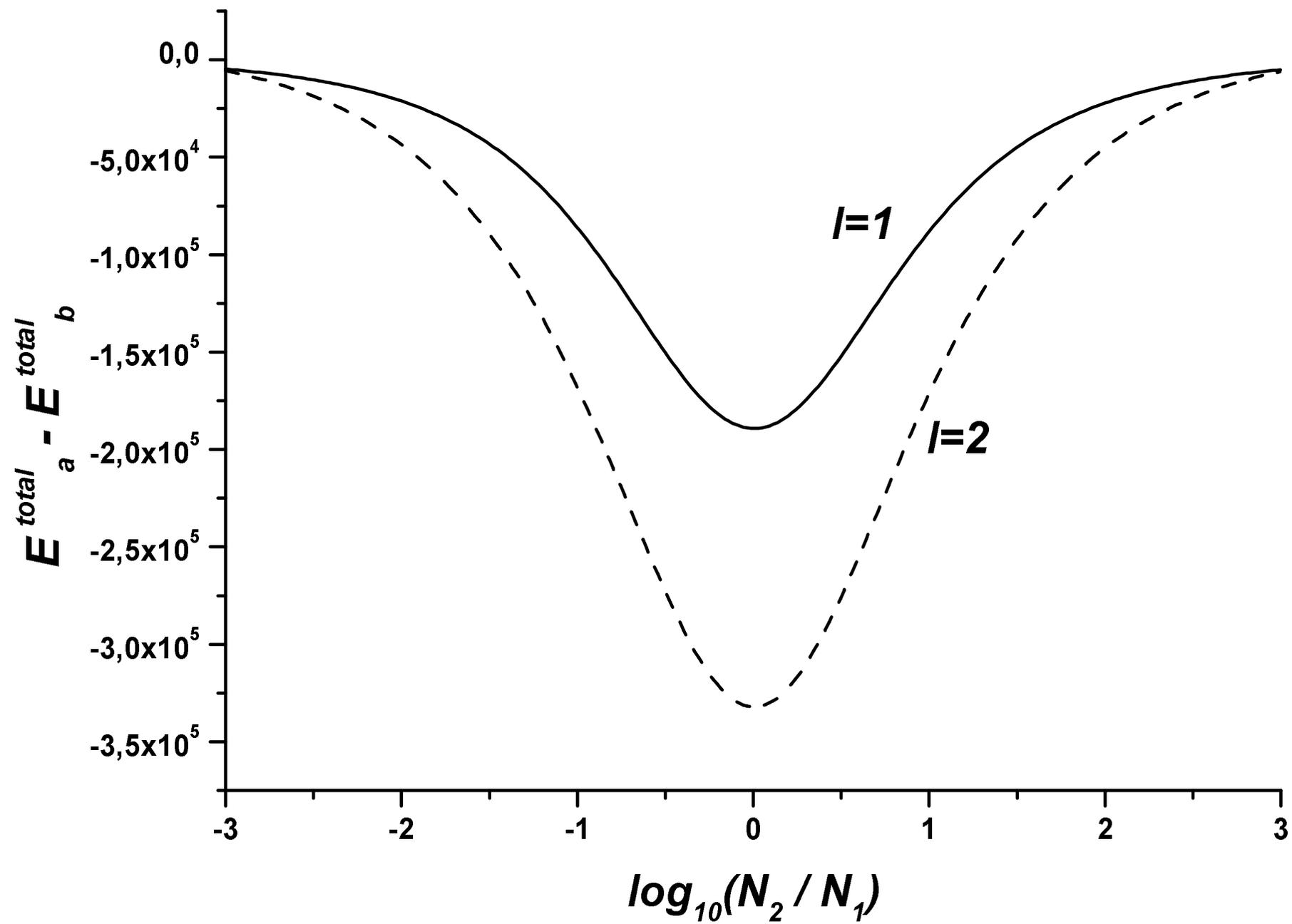
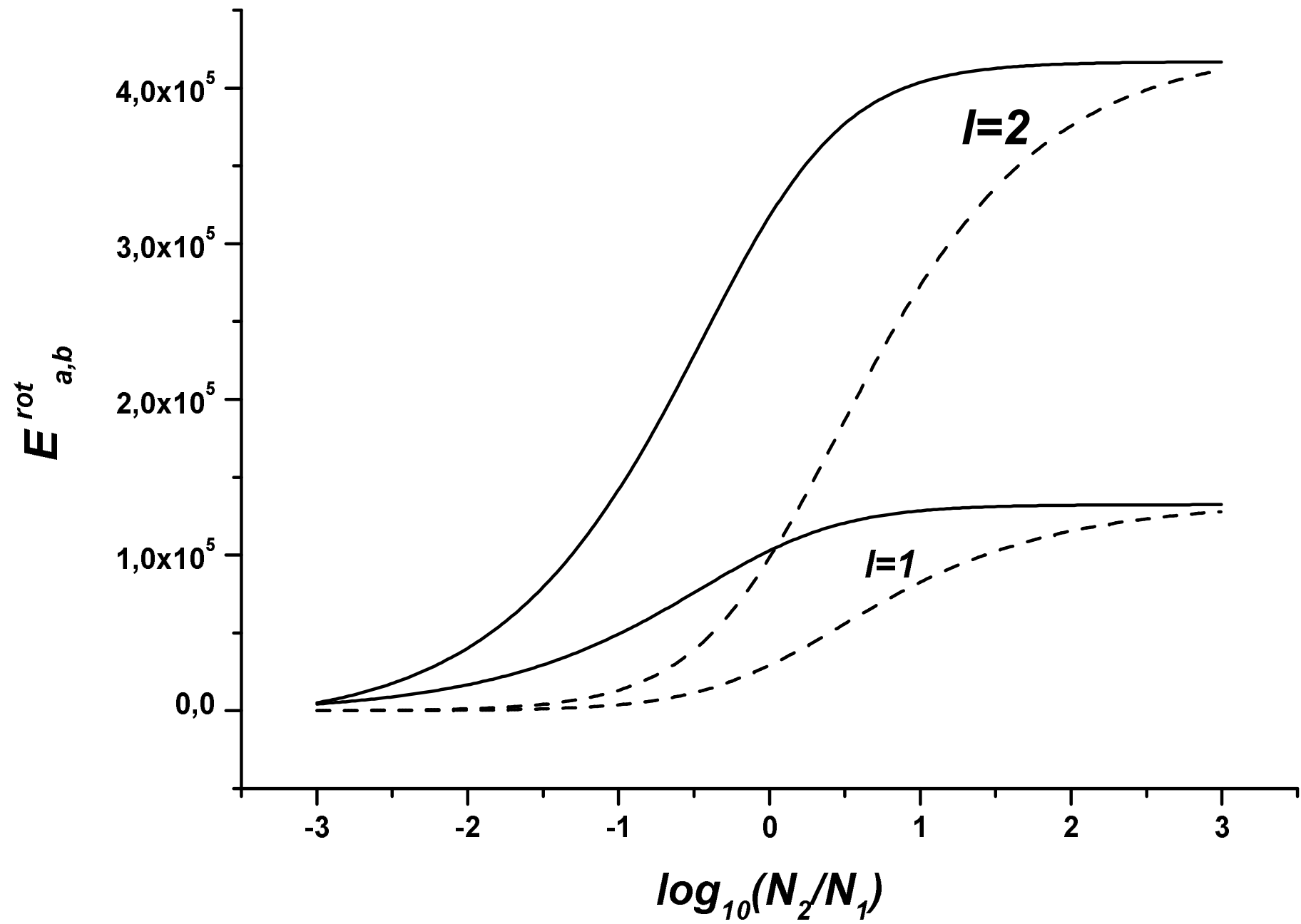
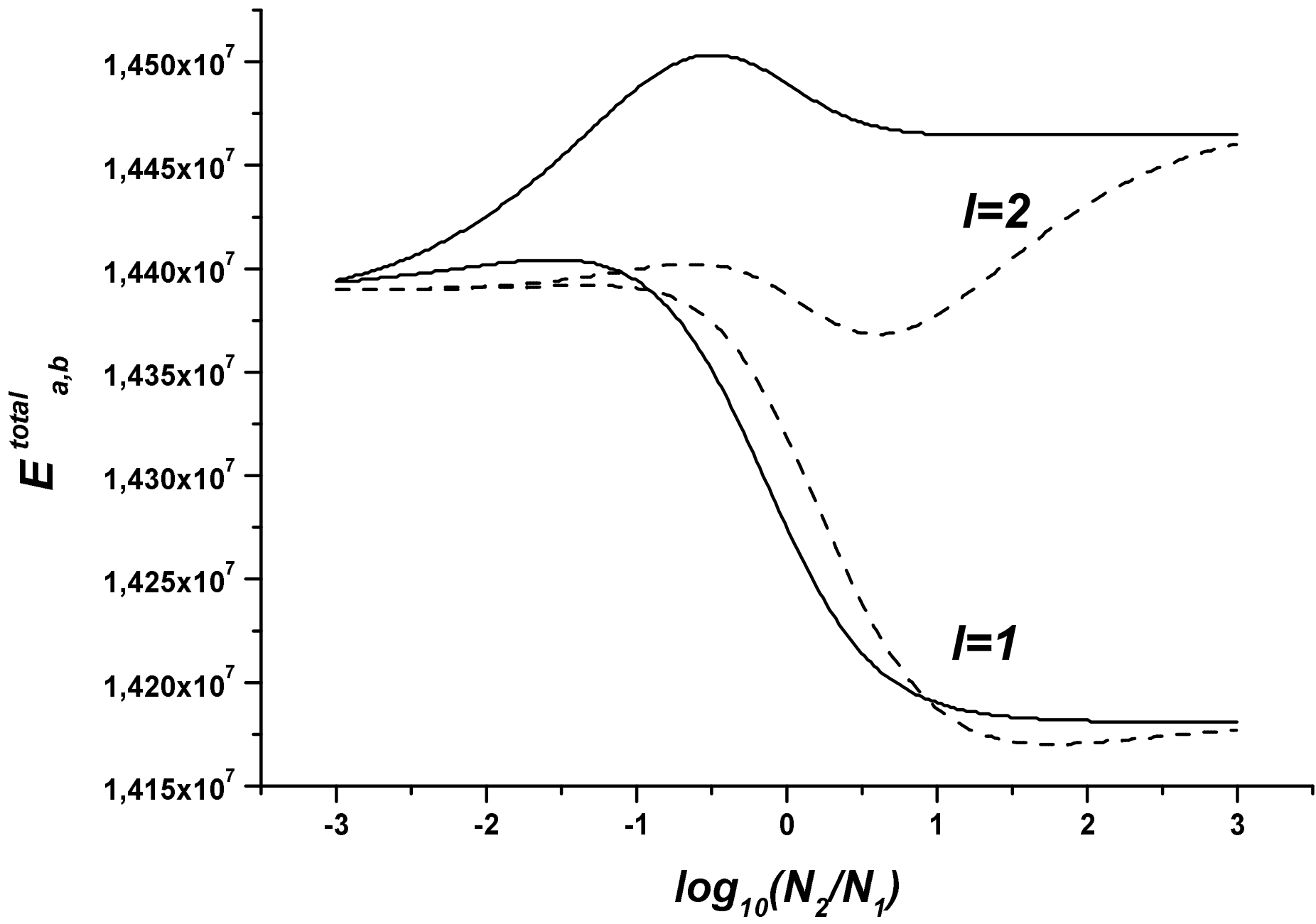
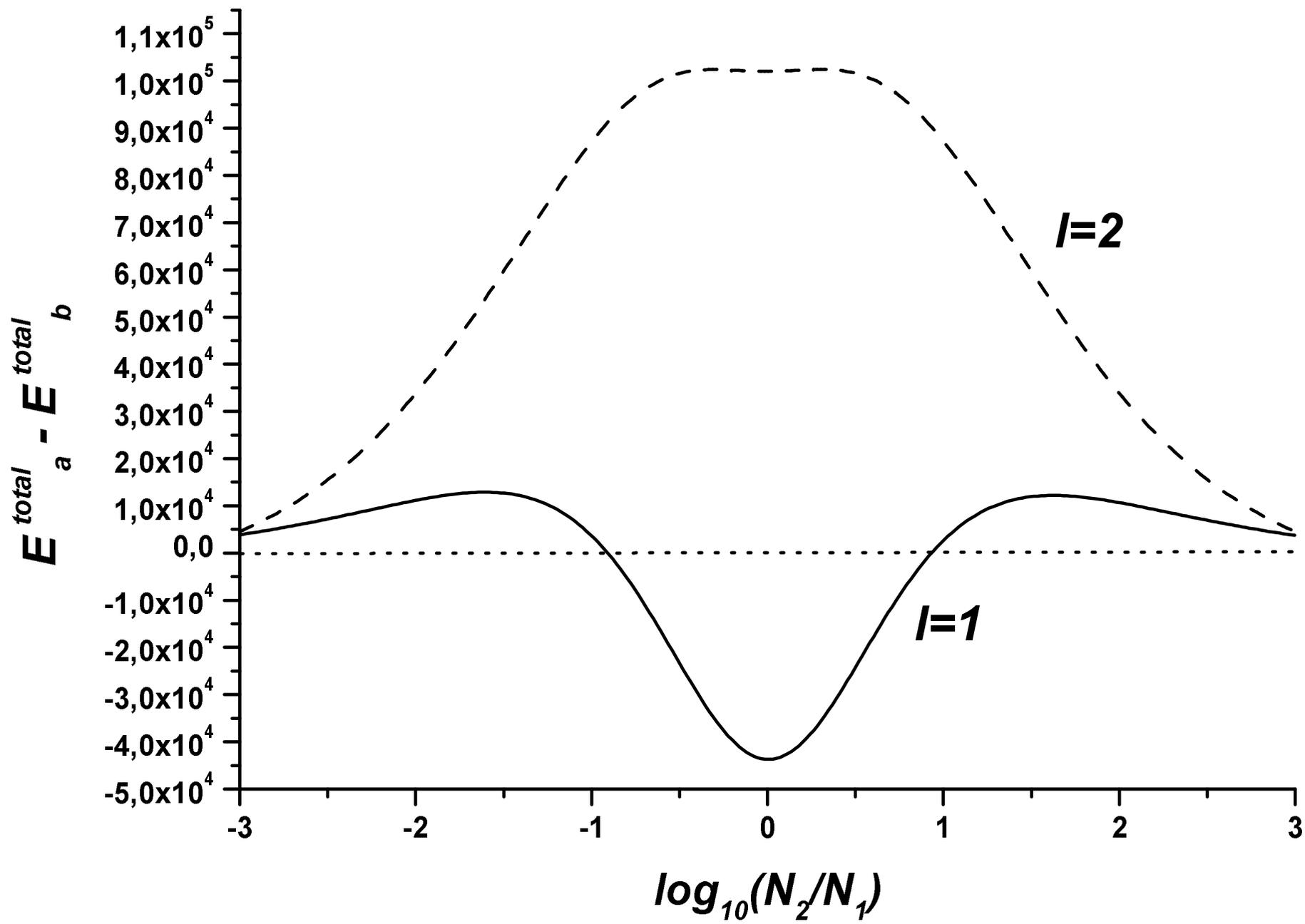
\end{document}